\documentclass[12pt]{article}

\usepackage[totalheight = 23cm, totalwidth = 17cm]{geometry}
\usepackage{amssymb,amsmath,amsfonts,amsbsy,graphicx,bm}

\usepackage{color,cancel,ulem}

\newcommand{\dis}[1]{\begin{equation}\begin{split}#1\end{split}\end{equation}}

\begin{document}

\begin{titlepage}

\begin{center}

{\LARGE \bf 
Von Neumann algebra description of inflationary cosmology
}

\vskip 1.0cm

{\large
Min-Seok Seo$^{a}$ 
}

\vskip 0.5cm

{\it
$^{a}$Department of Physics Education, Korea National University of Education,
\\ 
Cheongju 28173, Republic of Korea
}

\vskip 1.2cm

\end{center}

\begin{abstract}

  We study the von Neumann algebra description of the inflationary quasi-de Sitter (dS) space.
  Unlike   perfect dS space, quasi-dS space allows the nonzero energy flux across the horizon, which can be identified with the expectation value of the static time translation generator.
  Moreover, as a dS isometry associated with the static time translation is spontaneously broken, the fluctuation in time is accumulated, which induces the fluctuation in the energy flux.
  When the inflationary period is given by $(\epsilon_H H)^{-1}$ where $\epsilon_H$ is the slow-roll parameter measuring the increasing rate of the Hubble radius, both the energy flux and its fluctuation diverge in the $G \to 0$ limit.
  Taking the fluctuation in the energy flux and that in the observer's energy into account, we argue that the inflationary quasi-dS space is described by Type II$_\infty$ algebra.
  As the entropy is not bounded from above, this is different from Type II$_1$ description of perfect dS space in which the entropy is maximized by the maximal entanglement.
  We also show that our result is consistent with the observation that the von Neumann entropy for the density matrix reflecting the fluctuations above is interpreted as the generalized entropy.

\end{abstract}

\end{titlepage}

\newpage

\section{Introduction}

 Whereas the spacetime geometry close to de Sitter (dS) well describes the primordial inflation and the current accelerating expansion of the universe, understanding its quantum nature is  challenging.
 Studies on the quantum field theory in the dS background tell us that  a static observer in dS space is surrounded by the (cosmological) horizon of radius $r_H=H^{-1}$ having thermodynamic properties characterized by the Gibbons-Hawking temperature $\beta^{-1}=H/(2\pi)$ and the entropy $S_{\rm GH}=A/(4G)=\pi M_{\rm Pl}^2/H^2$, where $A$ is the horizon area given by $4\pi  r_H^2$ and $M_{\rm Pl}^2$ is defined as $G^{-1}$ \cite{Gibbons:1977mu}. 
 This is quite similar to the black hole as seen from far outside the horizon but the geometric structure of dS space different from that of the black hole gives rise to  several ambiguities.
 That is, unlike the black hole horizon, the boundary of the physically well defined compact object,  the cosmological horizon in dS space is observer dependent.
 Moreover, it is not clear that the dS entropy given by the finite number  counts the number of degrees of freedom of the region beyond the horizon which is not compact.
 
  Such ambiguities are expected to be fixed by the more complete   description of the thermodynamic behavior in quantum gravity.
  As an attempt to find it, it was recently proposed  that the entanglement needs to be described in terms of the algebra of local observables, rather than the tensor product of two  Hilbert spaces defined on two separated regions (for reviews, see, e.g., \cite{Witten:2018zxz, Witten:2021jzq}). 
  Here  the algebra of local observables ${\cal A}$ called the von Neumann algebra is required to satisfy the following conditions (see Section II. F of \cite{Witten:2018zxz}):
\begin{itemize}
\item Any element $a$ in ${\cal A}$ is bounded: this means that the eigenvalues $\alpha$ of $a$ are bounded, i.e., $|\alpha|<\infty$.
Then given the vector $|\psi\rangle$ in the Hilbert space consisting of the normalized vectors, $a|\psi\rangle$ is also normalizable, so belongs to the Hilbert space. 
More concretely, expanding $|\psi\rangle$ in terms of the eigenvectors $|\alpha\rangle$ of $a$ as $|\psi\rangle=\sum_\alpha c_\alpha|\alpha\rangle$ with $\langle\psi |\psi\rangle=\sum_\alpha |c_\alpha|^2<\infty$, the norm of $a|\psi\rangle=\sum_\alpha \alpha c_\alpha|\alpha\rangle$ given by $\sum_\alpha |\alpha|^2 |c_\alpha|^2$ is also finite provided $|\alpha|<\infty$.
\item The algebra is closed under the Hermitian conjugation : Typically, ${\cal A}$ consists of the localized operators called the `smeared fields', which is written in the form of $\phi_f=\int d^Dx f(x)\phi(x)$.
Here $f(x)$ is supported in the small region in which ${\cal A}$ is defined.
Then its complex conjugation $\phi_f^\dagger(x)=\int d^Dx f(x)^* \phi^\dagger(x)$ is also well defined and belongs to ${\cal A}$ as well.
\item The algebra is closed under the weak limit : if we consider the sequence of operators $a_1, a_2,\cdots \in {\cal A}$ then any matrix element $\langle \psi|a_n |\chi\rangle$ converges to $\langle \psi|a|\chi\rangle$ for some $a\in {\cal A}$ in the $n\to \infty$ limit.
From this, we can identify the observables with the finite (but the large number of) observations within the controllable error.
\end{itemize}  
  
  When the multiple of the identity   is the only allowed center, a subset of ${\cal A}$ consisting of operators that commute with all elements in ${\cal A}$, the von Neumann algebra is called a factor.
 Then the Hilbert space is constructed by acting the local  observables on the cyclic and separating state $|\Psi\rangle$. 
 Here the state $|\Psi\rangle$ is said to be cyclic for ${\cal A}$ if the states $a|\Psi\rangle$ for $a \in {\cal A}$ are dense in the Hilbert space, i.e.,  only the zero vector is orthogonal to all states in the form of $a|\Psi\rangle$.
Meanwhile, $|\Psi\rangle$ is called separating if $a=0$ is the only local observables in ${\cal A}$ satisfying $a|\Psi\rangle=0$.

 When we merely consider the quantized fluctuation   around the fixed background in the $G\to 0$ (or equivalently, $M_{\rm Pl}\to\infty$) limit, the algebra typically belongs to Type III, in which neither the pure state nor the entropy is well defined \cite{Araki:1964, Leutheusser:2021qhd, Leutheusser:2021frk, Gomez:2022eui}.
 \footnote{Formally, the state is called pure if the function $F_\Psi(a)=\langle \Psi | a|\Psi\rangle$ cannot be written in the form of $p_1 F_{\Phi_1}(a)+p_2 F_{\Phi_2}(a)$ with $p_{1,2}>0$, which means that decoherence does not take place thus interference effects appear.
 When the state is not pure, it is called mixed.
 Indeed, $F_\Psi(a)$ is used to define the density matrix through the trace, the linear functional of operators  satisfying  the commutative property and the positivity.
 If the finite trace is not defined, the divergent entropy   is not renormalized.
 For more complete discussion, see reviews \cite{Witten:2018zxz, Witten:2021jzq}.}
 By taking dynamical gravity into account through the ${\cal O}(G)$ corrections and treating diffeomorphism invariance as gauge redundancy or constraint, the algebra becomes Type II factor, in which the entropy as a finite, renormalized quantity can be defined.
 For the black hole, $H_{L/R}-M$, the deviation of the ADM Hamiltonian of the left/right patch around the ADM mass $M=r_s/(2G)$ ($r_s$ is the horizon radius) can take any real value in the $G\to 0$ limit as $M$ becomes divergent.
 Then the black hole thermodynamics is described by Type II$_\infty$ algebra \cite{Witten:2021unn, Chandrasekaran:2022eqq}, in which only a subset of observables has the finite trace hence the entropy is not bounded from above.
 In contrast, in dS space, there is no boundary at infinity.
 Instead, the static patch is bounded by the  horizon which is in thermal equilibrium with the Gibbons-Hawking radiation, resulting in the vanishing energy flux across the horizon.
 As we will see, this implies   the absence of the operator analogous to $H_{L/R}-M$ in the black hole, hence   the algebraic description of dS space is different from that of the black hole.
  In \cite{Chandrasekaran:2022cip}, it was found that when    the  static observer  has the positive energy as a random variable,  a system of dS space and the observer is well described by Type II$_1$ algebra, in which the trace of any bounded operator is finite. 
 As a result, the entropy has an upper bound, which is saturated for the maximally entangled state. 
 
 Meanwhile, in the inflationary  era,  $H$ is no longer a constant but a slowly varying function of the flat time coordinate $t$.
 Then some of  dS isometries are slightly broken and the  spacetime geometry is given by quasi-dS space.
  In this case,  the deviation of the background from perfect dS space   is parametrized by the slow-roll parameter $\epsilon_H$.
  When the inflation is driven by the vacuum energy of the  inflaton $\phi(t)$, a homogeneously evolving scalar field, $\epsilon_H$ is proportional to $\dot{\phi}^2$ :
  \dis{\epsilon_H \equiv \dot{r_H} = -\frac{\dot{H}}{H^2}=\frac{4\pi \dot{\phi^2}}{M_{\rm Pl}^2 H^2}=\frac{4\pi G \dot{\phi^2}}{H^2},\label{eq:epsilonH}}
 where dot denotes the derivative with respect to $t$. 
 In perfect dS case,  equations of motion are solved by the constant $H$ and $\dot{\phi}=0$, giving $\epsilon_H=0$.  
 On the other hand, even if $\dot{\phi}^2$ does not vanish, we can suppress $\epsilon_H$ close to zero by taking the $G\to 0$ limit, which we will focus on throughout this work.
 In any case, as $\epsilon_H \to 0$, the broken dS isometries are restored, implying the existence of the approximate timelike Killing vector associated with the static time coordinate $t_s$. 
 At the same time, the time scale $(\epsilon_H H)^{-1}$ after which $H$ is no longer   approximated as a constant becomes infinity. 
  We will explicitly show that when we take this time scale to be the inflationary period,  the energy flux across the horizon and its fluctuation become divergent in the $G\to 0$ limit. 
 Moreover, the energy flux across the horizon is  interpreted as the expectation value of the static time translation generator, and   its fluctuation  is driven by the fluctuation in time, hence that in the value of $H$ at the end of inflation.
 Then we find that unlike perfect dS space, the inflationary quasi-dS space is described by Type II$_\infty$ algebra  rather than Type II$_1$ algebra.

 The organization of this article is as follows.
 In Section \ref{sec:hordyn}, we  describe how the change in the horizon area induced by the slow-roll gives rise to the nonzero   energy flux across the horizon, which is identified with the expectation value of the static time translation generator.
   In Section \ref{sec:II1dS}, we observe the modification of the von Neumann algebra description of the inflationary quasi-dS space   from that of perfect dS space when we take the nonzero energy flux across the horizon into account.
  After claiming that quasi-dS space is well described by Type II$_\infty$ algebra, we provide the expression for the von Neumann entropy of the static patch in Section \ref{sec:rhoandS}.
  In Section \ref{sec:extquasidS}, we relate this with the change in the horizon area considered in Section \ref{sec:hordyn} to complete our  argument.  
  Then we conclude with a brief comment about the possibility that the inflationary quasi-dS space is described by Type II$_1$ algebra.
  This can happen  when the inflationary period is much shorter than $(\epsilon_H H)^{-1}$ as recently conjectured in the swampland program.
 In Appendix \ref{app:dSdic}, we summarize  various coordinates on dS space which are used throughout the discussion.
 In Appendix \ref{app:derden}, details of the density matrix considered in Section \ref{sec:rhoandS} are given.

\section{Horizon dynamics of quasi-dS space }
\label{sec:hordyn}

 \begin{figure}[!t]
  \begin{center}
   \includegraphics[width=0.45\textwidth]{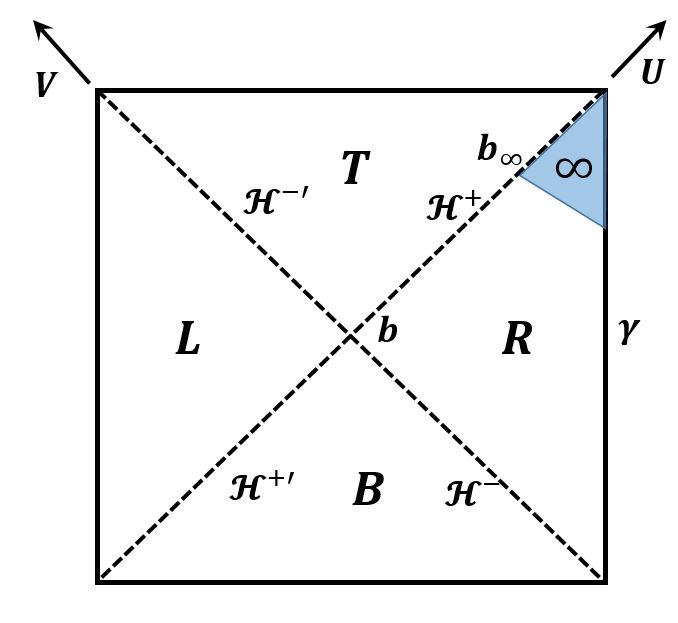}
  \end{center}
 \caption{Penrose diagram of dS space.
 Over the whole period of inflation, a static observer travelling along the timelike trajectory $\gamma$ is causally accessible to the static patch $R$, the region bounded by the future horizon ${\cal H}^+$ and the past horizon ${\cal H}^-$.
 Two horizons intersect at the bifurcation cut $b$.
 At a time close to the end of inflation, the static observer is causally connected to the region denoted by $\infty$, the right of the future horizon cut    $b_\infty$. 
 In the inflationary era, the universe covers the region  $R \cup T$, the past boundary  ${\cal H}^- \cup {\cal H}^{-\prime}$ of which corresponds to the initial singularity. 
 The complement static patch is denoted by $L$.
 The increasing directions of Kruskal-Szekeres coordinates $U$ and $V$ are also depicted.
  }
\label{Fig:dSpenrose}
\end{figure}

In this section, we estimate the energy flux across the future horizon ${\cal H}^+$ during the inflationary period, which relates the change in the horizon area to the static time translation generator.
We assume that there is no  energy flux across the initial singularity, ${\cal H}^- \cup {\cal H}^{-\prime}$ in Figure \ref{Fig:dSpenrose},  the past boundary  of the region $T\cup R$ covered by the flat coordinates.
Since the past horizon ${\cal H}^-$ belongs to the initial singularity, we do not consider the energy flux across ${\cal H}^-$.

 The energy-momentum tensor of the inflaton field $\phi$  with the canonical kinetic term is given by 
\dis{T_{\mu\nu}=\nabla_\mu \phi \nabla_\nu \phi-\Big(\frac12(\nabla\phi)^2+V(\phi)\Big)g_{\mu\nu}.}
 Since spacetime during inflation is homogeneous and isotropic at large scale, we expect that the equations of motion are solved by $\phi(t)$ which depends   only on the flat time coordinate $t$ (the metric of dS space in the flat coordinates $(t, r)$ can be found in \eqref{eq:flatmetric}).  
Since $\dot{\phi}$ becomes zero in the perfect dS limit ($H=$constant), it measures the deviation of the background from dS space, which is evident from \eqref{eq:epsilonH}, i.e.,  $\epsilon_H \propto \dot{\phi}^2$.
 So far as $\epsilon_H$ is very tiny, one can find the approximate dS isometries, which allow  an approximate timelike Killing vector along the direction of the static time coordinate $t_s$, $k^a =(\partial_{t_s})^a =(\partial_t - H r\partial_r)^a$ (the metric of dS space in the static coordinates $(t_s, r_s)$ can be found in \eqref{eq:staticmetric}). 
The component of the energy-momentum tensor  associated with the $t_s$ direction is written as  
\dis{T_{t_st_s}=\dot{\phi}^2+(1-H^2r_s^2) \Big(-\frac12\dot{\phi}^2+V(\phi)\Big),}
hence $T_{t_st_s}=\dot{\phi}^2$ on the horizon $r_s=r_H=H^{-1}$.
Furthermore, the relation  $\partial_{r_s}\phi=(\partial t/\partial r_s)\dot{\phi}=-[(Hr_s)/(1-H^2r_s^2)]\dot{\phi}$ (see \eqref{eq:statflat2}) gives
\dis{T_{t_s r_s}=-\frac{Hr_s}{1-H^2r_s^2}\dot{\phi}^2.}
For a more straightforward interpretation of this, we consider $T_{t_sr_*}$ by converting $r_s$ into the tortoise coordinate $r_*$ defined in \eqref{eq:tortoise}.
From this, we can define `luminosity', the energy flux across the surface of constant $r_s$ by ${\cal L}=-4\pi r_s^2 T_{t_sr_*}$ \cite{Schutz:2009}.
Since 
\dis{T_{t_s r_*}= -H r_s \dot{\phi}^2}
becomes $T_{t_s r_*}= -\dot{\phi}^2$ on the horizon,  the luminosity on the horizon is given by ${\cal  L}=4\pi H^{-2}\dot{\phi}^2=\epsilon_H M_{\rm Pl}^2$.
 Meanwhile, the Kruskal-Szekeres coordinates $U$ and $V$, in terms of which the metric is written as \eqref{eq:UVmetric}, are natural affine parameters on ${\cal H}^+$ and ${\cal H}^-$, respectively.
 The energy-momentum tensor components in the Eddington-Finkelstein coordinates ($t, r_*$) and those in the Kruskal-Szekeres coordinates ($U, V$) are related as
 \dis{&T_{t_st_s}=H^2U^2 T_{UU}+H^2V^2T_{VV}-2 H^2 UV T_{UV},
 \\
 &T_{t_s r_*}=-H^2U^2T_{UU}+H^2V^2T_{VV}.}
Then the simple relations 
 \dis{T_{UU}=\frac{T_{t_st_s}}{H^2U^2}=-\frac{T_{t_sr_*}}{H^2U^2} = \frac{\dot{\phi}^2}{H^2U^2}\label{eq:Trelat}}
 are satisfied on ${\cal H}^+$ ($V=0$).
 
 The  energy-momentum tensor components on ${\cal H}^+$ are used to find the first law of thermodynamics, which relates the energy flux across the horizon to the change in the horizon area \cite{Frolov:2002va, GalvezGhersi:2011tx}.
 In the perfect dS limit, we can use the timelike Killing vector $k^a=(\partial_{t_s})^a=H(U\partial_U-V\partial_V)$ to find the conserved  current
  \dis{J^a=-T^a_{~b}k^b.}
Since $V=0=$(constant) on ${\cal H}^+$, relations $k^a=HU\partial_U$ and $dV=0$ thus $g^{ab}(dV)_a (dV)_b=0$  are satisfied, implying that the vector $(\partial_U)^a$ which is proportional to $g^{ab}(dV)_b$ is normal as well as tangential to ${\cal H}^+$.
Then  ${\cal H}^+$ corresponds to the Killing horizon and the energy flux across ${\cal H}^+$ is given by 
\dis{\Delta E &=-\int_{{\cal H}^+}d\Sigma_a  J^a = \int_{{\cal H}^+} d\Omega dU\sqrt{\gamma} T_{Ut_s}
=\int_{{\cal H}^+} d\Omega dU \sqrt{\gamma} (HU)T_{UU},}
where $d\Sigma_a$ is the volume element on ${\cal H}^+$ and $\sqrt{\gamma}=r_H^2=H^{-2}$.
As can be inferred from the relation $k^a=HU\partial_U$ on ${\cal H}^+$ and $T_{Ut_s}$ in the integrand, $\Delta E$ is interpreted as the   static time translation generator on ${\cal H}^+$.
More precisely, consider the semiclassical approximation in which the $\hbar \to 0$ limit is taken and  the quantum fluctuations around the mean values of the operators are sufficiently small.
\footnote{But still, the solution can be fluctuated by the statistical uncertainty or the `classical' fluctuation.
This is generated by the decoherence, the loss of the interference effect through the interaction of the system with the   environment, the region the observer is ignorant of. 
See discussion in the paragraph containing \eqref{eq:flucphi} and references therein.}
Denoting the quantum state of the inflationary universe by $|\Phi\rangle$ and the operator generating the static time translation on the `boundary' ${\cal H}^+$ of the static patch $R$ by $H_R$,   the solutions to the equations of motion ${\phi}(t)$ and $T_{\mu\nu}$ can be regarded as $\langle \Phi|\phi |\Phi\rangle$ and  $\langle \Phi|T_{\mu\nu}|\Phi\rangle$, respectively.
Then $\Delta E$ is identified with $\langle \Phi|H_R|\Phi\rangle$.
Indeed, the horizon is regarded as a boundary of the  static patch.
Whereas the bulk Hamiltonian as a constraint associated with the static time translation vanishes which is evident in the ADM formalism, the boundary Hamiltonian defined on the horizon is nonzero, and plays the role of the generator of the static time translation.
As we will see, $\Delta E$ is proportional to the horizon area, and  as argued in    Section 2.4 and Section 3 of \cite{Chandrasekaran:2022eqq} (and references therein), it is canonically conjugate to the `boost' (static time translation) so identified with the static time translation generator.

Meanwhile, since $r_s=r_H=H^{-1}$ is almost constant   on ${\cal H}^+$, the  relations $U=H^{-1}e^{H(t_s-r_*)}$ (see \eqref{eq:UVDef}) and $t_s=t-\frac{1}{2H}\log\big(1-H^2 r_s^2 \big)$ (see \eqref{eq:statflat1}) indicate that  $dU=HU dt$ is satisfied on ${\cal H}^+$. 
Then from \eqref{eq:Trelat} one finds
\dis{\langle \Phi|H_R|\Phi\rangle=\Delta E
= \int  d t   \frac{4\pi}{H^2}\dot{\phi}^2=\int dt \epsilon_H M_{\rm Pl}^2,
\label{eq:deltaE}}
where the integrand is nothing more than the luminosity ${\cal L}$ and the range of $t$ integration is taken   to be the inflationary period during which $H$ is almost   constant.
In addition, we  assume  $|\dot{\epsilon_H}/(\epsilon_H H)|\ll 1$ such that $\epsilon_H$ does not vary much during the inflationary period.
Since the value of $H$ considerably deviates from the initial value after $\Delta t={\cal O}(1)\times(\epsilon_H H)^{-1}$,
we may take $t \in (-c(2\epsilon_H H)^{-1}, c(2\epsilon_H H)^{-1})$ with $c$ being some constant smaller than $1$ (not to  spoil the perturbative expansion with respect to $\epsilon_H$), which  becomes $t \in (-\infty, +\infty)$ in the perfect dS limit $\epsilon_H \to 0$. 
Then $\langle \Phi|H_R|\Phi\rangle$ is estimated as 
\dis{\langle \Phi|H_R|\Phi\rangle \simeq \frac{c}{\epsilon_H H}\times  \epsilon_H M_{\rm Pl}^2 = c \frac{M_{\rm Pl}^2}{H}, \label{eq:deltaEest}}
i.e., $M_{\rm Pl}^2/H$ up to ${\cal O}(1)$ (but smaller than $1$) coefficient, showing that $\langle \Phi|H_R|\Phi\rangle$ is insensitive to $\epsilon_H$, or equivalently, $\dot{\phi}^2$ at leading order.

 The backreaction of the energy flux across the horizon leads to the deformation of the geometry   parametrized by expansion, shear, and rotation.
 When the  background is close to dS space, the horizon can be approximated as a Killing horizon, where all the three parameters vanish at leading order.
 Then the Raychaudhuri equation for the expansion $\Theta = A^{-1}dA/dU$ which describes the change in the horizon area $A=4\pi r_H^2=4\pi H^{-2}$ is approximated as
 \dis{\frac{d\Theta}{dU} \simeq -\frac{8\pi}{M_{\rm Pl}^2}T_{ab}(\partial_U)^a (\partial_U)^b = -\frac{8\pi}{M_{\rm Pl}^2}T_{UU},}
 from which we can replace $T_{UU}$ by $-[M_{\rm Pl}^2/(8\pi)]d\Theta/dU$.
Putting this into $\langle \Phi|H_R|\Phi\rangle$, we obtain
\dis{\langle \Phi|H_R|\Phi\rangle&=-\frac{M_{\rm Pl}^2}{8\pi}\int d\Omega dU\sqrt{\gamma} (HU)\frac{d\Theta}{dU}
\\
&=-\frac{M_{\rm Pl}^2}{8\pi}\int d\Omega\Big[\sqrt{\gamma}HU \Theta\Big|_{U=0}^{U\simeq \infty}-\int dU\Big(\sqrt{\gamma}H+H U\frac{d\sqrt{\gamma}}{dU}+\sqrt{\gamma}U\frac{dH}{dU}\Big)\Theta\Big].\label{eq:deltaEint}}
Noting that $\partial_U=(HU)^{-1}\partial_{t_s}$ and $dt_s=dt$ on ${\cal H}^+$, one  finds that $dH/dU=-\epsilon_H H/U$, $d\sqrt{\gamma}/dU=2\epsilon_H/(H^2U)$  and $\Theta=2\epsilon_H/U$. 
Then the last two terms  in \eqref{eq:deltaEint}  are ${\cal O}(\epsilon_H^2 \Delta t)$.
For the first surface term, since $\sqrt{\gamma}HU\Theta$ is ${\cal O}(\epsilon_H)$,  the variation of $\sqrt{\gamma}HU\Theta$ over ${\cal H}^+$ is ${\cal O}(\epsilon_H^2 \Delta t)$.
Therefore, the second term in \eqref{eq:deltaEint} gives the  leading contribution to $\langle \Phi|H_R|\Phi\rangle$ of ${\cal O}(\epsilon_H \Delta t)$ : 
\dis{\langle \Phi|H_R|\Phi\rangle&=\frac{M_{\rm Pl}^2}{8\pi}\int  dU H \int d\Omega \sqrt{\gamma}  \Theta +{\cal O}(\epsilon_H^2 \Delta t)=
\frac{M_{\rm Pl}^2}{8\pi}\int dU H \frac{dA}{dU} +{\cal O}(\epsilon_H^2 \Delta t)
\\
&=\frac{M_{\rm Pl}^2}{8\pi}\int d t H \frac{dA}{dt} +{\cal O}(\epsilon_H^2 \Delta t) = \int dt \frac{H}{2\pi}\frac{d}{dt}\Big(\frac{A}{4G}\Big)+{\cal O}(\epsilon_H^2\Delta t).}
Since the Gibbons-Hawking temperature is given by $H/(2\pi)$, the integrand can be written in the form of $T\Delta S$, which is consistent with the first law of thermodynamics. 
We also note that the explicit calculation of the integrand reproduces \eqref{eq:deltaE}.
 To see the physical meaning of $\langle \Phi|H_R|\Phi\rangle=\Delta E$ more clear, we recall that the static time translation is a diffeomorphism, the gauge invariance of gravity, hence it acts as a constraint on the dynamics of quantum gravity.
 As a result, just like the ADM mass of the black hole, the associated  charge gets contribution from the surface integral on the boundary (${\cal H}^+$ for dS space) only, which is given by $\beta^{-1}S=M_{\rm Pl}^2/(2 H)=r_H/(2G)$ \cite{Frodden:2011eb}.
 This is supported by the fact that the energy inside the horizon in the perfect dS limit is estimated as $M_{\rm Pl}^2/(2H)$, which is obtained by multiplying  the energy density during inflation $T_{tt}\simeq [3/(8\pi)]M_{\rm Pl}^2 H^2$ by the volume inside the horizon   $(4\pi/3)H^{-3}$.
\footnote{The radius of the horizon in the flat coordinates $H^{-1}e^{-Ht}$ gives $V=(4\pi/3) e^{3Ht}\times (H^{-1}e^{-Ht})^3$ where the factor $e^{3Ht}$ comes from $\sqrt{-g}$ restricted to the spatial directions.}
If we regard the slow-roll as the adiabatic process, $H$ slowly decreases in time, then the `ADM mass' just after the end of inflation  $M_{\rm Pl}^2/(2H)+\Delta E$ is identified with $M_{\rm Pl}^2/(2H_f)$ where $H_f$ is the value of $H$ at that time.
As $H_f$ significantly deviates from the initial $H$, we expect that $\Delta E$ is at most given by the same order as $M_{\rm Pl}^2/(2H)$.

 It is remarkable that $\langle \Phi|H_R|\Phi\rangle$ becomes divergent in the $G\to 0$, or equivalently, $ M_{\rm Pl}^2 \to \infty$  limit.
Indeed, whereas the perfect dS limit $\epsilon_H \to 0$    is trivially obtained by taking $\dot{\phi}\to 0$, we can also reach the same limit by taking $G\to 0$ even if  $\dot{\phi}^2$ is kept finite, as can be noticed from \eqref{eq:epsilonH}.
In this case, as $\epsilon_H$ almost vanishes, the spacetime geometry can be well approximated by the perfect dS space.
Moreover, even though the effects of $\epsilon_H$ become  negligible,  the ratio $\epsilon_H/S_{\rm GH}=4\pi^2\dot{\phi}^2/H^4$, which is independent of $G$, can be kept nonvanishing.
This is quite similar to what was assumed to find the AdS/CFT correspondence : the string coupling $g_s$ is taken to vanish to decouple string interactions, but the `t Hooft coupling $g_{\rm YM}^2 N \sim g_s  N$ is kept fixed \cite{Maldacena:1997re}.
Therefore, by keeping $\epsilon_H/S_{\rm GH}>1$, i.e., $\epsilon_H > H^2/M_{\rm Pl}^2$, we can still forbid  eternal inflation, and the formulation based on the semiclassical approximation is reliable against large quantum fluctuations or the non-perturbative effects \cite{Arkani-Hamed:2007ryv}.
Indeed, recent swampland conjectures concerning the instability of dS space claimed the lower bound on the potential slope \cite{Obied:2018sgi, Andriot:2018mav, Garg:2018reu, Ooguri:2018wrx}, which forbids eternal inflation (see, e.g.,  \cite{Kinney:2018kew, Brahma:2019iyy, Rudelius:2019cfh, Rudelius:2019cfh}).
But at the same time, as implied by the nonzero energy flux, the horizon is no longer in thermal equilibrium with the Gibbons-Hawking radiation. 
Then the state $|\Phi\rangle$ breaks the dS isometry  by allowing the nonzero $\langle \Phi|H_R|\Phi\rangle$, instead of being annihilated by $H_R$, just like   the Unruh state describing the evaporating black hole \cite{Unruh:1976db} (see also \cite{Aalsma:2019rpt, Gong:2020mbn} for recent discussions).
This can be contrasted with  perfect dS space ($\dot{\phi}=0$), in which the horizon is in thermal equilibrium hence the energy flux across the horizon vanishes.
Then the quantum state $|\Psi\rangle$ for  perfect dS space respects the dS isometry.
This   is called the Bunch-Davies state \cite{Chernikov:1968zm, Bunch:1978yq}, the dS analogy of the Hartle-Hawking state of the black hole \cite{Hartle:1976tp}. 

 Before moving onto the fluctuation, we  comment on our assumption $|\dot{\epsilon_H}/(\epsilon_H H)|\ll 1$.
 Since 
\dis{\frac{\dot{\epsilon_H}}{\epsilon_H H}=\frac{\ddot H}{H {\dot H}}-2\frac{\dot H}{H^2}=2(-\eta_H+\epsilon_H), }
where $\eta_H \equiv -{\ddot{H}}/(2 {\dot H}H)$ is another slow-roll parameter, this assumption indicates that $\epsilon_H \simeq \eta_H$.
On the other hand, there is a priori no  reason that $\epsilon_H$ and $\eta_H$ are similar in size : we just require that these two parameters are sufficiently smaller than $1$.
To see the role of $\eta_H$,  we consider the perturbative expansion of $H$ around some pivotal value $H_0$, say, the initial value of $H$, 
\dis{H(t)&=H_0 +\dot{H}_0 \Delta t +\frac12 \ddot{H}_0 \Delta t^2+\cdots
\\
&=H_0\big(1 -\epsilon_H (H_0\Delta t)+\eta_H \epsilon_H (H_0\Delta t)^2+\cdots\big),}
which is valid for $\epsilon_H, |\eta_H| <1$ and $H_0\Delta t <{\rm min}(1/\epsilon_H, 1/|\eta_H|)$.
In particular, $\Delta t$ is at most $c\times  {\rm min}(1/\epsilon_H, 1/|\eta_H|)$ with $c$ being a constant smaller than $1$, which can be employed as the inflationary period.
Denoting the value of $H(t)$ at the end of inflation, i.e., $H_0 \Delta t \simeq c\times  {\rm min}(1/\epsilon_H, 1/|\eta_H|)$, by $H_f$, and using the definition of $\epsilon_H\equiv -{\dot H}/H^2$, one finds that $\Delta E$ can be written as
\dis{\Delta E&=\int dt \epsilon_H M_{\rm Pl}^2 =-\int\frac{dH}{H^2 }M_{\rm Pl}^2=M_{\rm Pl}^2\Big(\frac{1}{H_f}-\frac{1}{H_0}\Big)
\\
&=\epsilon_H \Delta t +(\epsilon_H^2-\epsilon_H\eta_H)H_0 \Delta t^2+\cdots,}
which is positive.
Therefore, if $\epsilon_H \gtrsim |\eta_H|$, we take $H\Delta t \simeq c/\epsilon_H$, which leads to 
\dis{\Delta E=\frac{M_{\rm Pl}^2}{H_0} \Big(c+(\epsilon_H^2-\epsilon_H \eta_H)c^2+\cdots\Big),}
that is, with appropriately chosen $c<1$, $\Delta E\sim c M_{\rm Pl}^2/H$, as we considered so far.
In contrast, if $\epsilon_H < |\eta_H|$, we take $H\Delta t \simeq c/|\eta_H|$, giving
\dis{\Delta E=\frac{M_{\rm Pl}^2}{H_0} \Big(c\frac{\epsilon_H}{|\eta_H|}+\Big(\frac{\epsilon_H^2}{\eta_H^2}- \frac{\epsilon_H}{\eta_H}\Big)c^2+\cdots\Big),
}
which is smaller than $M_{\rm Pl}^2/H$.
This bound is saturated when $\epsilon_H$ becomes close to $ \eta_H$.
This shows that our estimation $\Delta E \sim M_{\rm Pl}^2/H$ up to a constant smaller than $1$ in fact corresponds to the maximal value of the increment in the ADM mass.
Treating $\epsilon_H$ and $\eta_H$ as independent parameters, one may regard the ratio $\epsilon_H/|\eta_H|$ to be another factor smaller than $1$ multiplied to $M_{\rm Pl}^2/H$.


 Our discussion so far is made in terms of the solutions to the classical equations of motion, which are regarded as the expectation values of the   operators with respect to $|\Phi\rangle$. 
 On the other hand, as a dS isometry associated with the static time translation is spontaneously broken by the quasi-dS background, the quantum fluctuation in $\phi$ combines with that in the trace of the spatial  metric \cite{Cheung:2007st, Weinberg:2008hq} (see also \cite{Prokopec:2010be, Gong:2016qpq}), forming the gauge invariant operator which excites   the curvature perturbation \cite{Mukhanov:1985rz, Sasaki:1986hm}.
 As the universe undergoes   accelerated expansion, the wavelength of the curvature perturbation is stretched beyond the horizon scale, after which the fluctuation can be treated as a classical one through the open system description called decoherence \cite{Hartle:1992as, Burgess:2006jn, Burgess:2014eoa, Nelson:2016kjm, Shandera:2017qkg, Gong:2019yyz}.
 In terms of the standard cosmological perturbation theory, this can be explained by the fact that the perturbation with wavelength larger than $H^{-1}$ no longer oscillates and may be treated as a frozen distribution of a {\it classical} field $\phi(t)$ 
\cite{Linde:1986fd}.
 This contributes to the accumulated uncertainty of the classical trajectory $\phi(t)$ during $\Delta t$ given by \cite{Linde:1986fd} (see also \cite{Vilenkin:1982wt, Linde:1982uu, Starobinsky:1982ee})
\dis{\langle \phi(t)^2 \rangle\equiv \delta \phi^2=\Big(\frac{H}{2\pi}\Big)^2 H\Delta t.\label{eq:flucphi}}
Noting   $\beta^{-1}=H/(2\pi)$,  the accumulated uncertainty  may be interpreted as  the thermal fluctuation,   which induces the fluctuation in $H_R$ estimated as 
\dis{\delta H_R =\Big(\frac{\partial}{\partial\beta}\langle \Phi|H_R|\Phi\rangle\Big)^{1/2}\simeq\frac{M_{\rm Pl}}{\sqrt{2\pi}},}
where the multiplication by $c^{1/2}$ is implicit.
We can reach the similar conclusion in the following way.
In perfect dS space, different constant $t$ (flat time) slices  are  physically equivalent  due to the isometry associated with the $t_s$ (static time) translation : the  translation of $t$ can be compensated by the scaling of $r$, leaving $r_s=r e^{Ht}$ (see \eqref{eq:statflat1}) hence the metric in the static coordinates \eqref{eq:staticmetric} unchanged.
In quasi-dS space, however, the time dependent classical  solution $\phi(t)$ plays the role of `clock' distinguishing the specific constant $t$ slice from others.
Then the fluctuation in $\phi(t)$ given by \eqref{eq:flucphi} leads to the fluctuation in time $\delta t$,
\footnote{This should not be confused with $\Delta t$, the time interval without fluctuation during which the fluctuation  in $\phi(t)$ is accumulated : $t$ is the time appearing in the background geometry, at each instant of which the fluctuation of $t$ given by $\delta t$ is accumulated by $\delta \phi$.
}
 thus that in $H$ :
\dis{|\delta H|= |\dot{H}\delta t|=\Big|\frac{\dot{H}}{\dot{\phi}}\delta \phi\Big|=\sqrt{\frac{\epsilon_H}{\pi}}H\frac{H}{M_{\rm Pl}}(H\Delta t)^{1/2}.}
From this, we can estimate the fluctuation in $\langle \Phi|H_R|\Phi\rangle$ over the inflationary period $\Delta t=c(\epsilon_H H)^{-1}$  to obtain
\dis{|\delta H_R|=c^{1/2}\frac{M_{\rm Pl}^2}{H^2}|\delta H|\simeq \frac{M_{\rm Pl}}{\sqrt{\pi}},}
which diverges in the $G \to 0$ limit.

 Since $\langle \Phi| H_R |\Phi\rangle  \sim {\cal O}(G^{-1})$ is divergent in the $G\to 0$ limit, one may define the `renormalized' static time translation generator by  $H'_{R}=H_{R}-\langle \Phi| H_R |\Phi\rangle$.
 But still, $\langle \Phi| (H'_R)^2 |\Phi\rangle =(\delta H_R)^2 \sim {\cal O}(G^{-1})$ is also divergent in the $G\to 0$ limit, so $H'_R|\Phi\rangle$ has a divergent norm and $H'_R$ is not well defined.
In order for the static time translation generator to be well defined, i.e., the expectation values of both the generator and its square to be finite in the $G\to 0$ limit,  we  need to extend the spacetime  to  the regions $L$ and $B$ in Figure \ref{Fig:dSpenrose} such that $|\Phi\rangle$ describes the whole quasi-dS manifold covering $R\cup T \cup L \cup B$, as suggested in Section 2 of \cite{Witten:2021unn}.
 Then we can introduce $H_L$, the static time translation generator in the complementary static patch $L$.
 Since $L$ is just a copy of $R$,  the energy flux across the future horizon ${\cal H}^{+\prime}$ is the same in size as that across ${\cal H}^{+}$.
But   the static time in  $L$ flows in the opposite   direction to that in $R$, so $\Delta E$ through  ${\cal H}^{+}$ (say, flowing from $R$ to $T$) has an opposite sign to that through ${\cal H}^{+\prime}$ (say, flowing from $L$ to $B$). 
 From this, we expect that $\Delta E$ on ${\cal H}^{+\prime}$ is identified with   $-\langle \Phi | H_L |\Phi\rangle$ and  the sum of $\Delta E$ on ${\cal H}^+$ and that on ${\cal H}^{+\prime}$ vanishes, giving $\langle \Phi | H_R |\Phi\rangle=\langle \Phi | H_L |\Phi\rangle$.  
 Now let us define  the total static time translation generator    by $H_0=H_R-H_L = H'_R-H'_L$.
 This acts on the thermofield double state $|{\rm TFD}\rangle=\sum_n c_n|E_n\rangle_R|E_n\rangle_L$ describing the entanglement between states living on ${\cal H}^+$ and ${\cal H}^{+\prime}$.\
 Since  both $H_0$ and $(H_0)^2$ annihilate  the thermofield double state ($H_0|{\rm TFD}\rangle =\sum_n c_n (E_n-E_n)|E_n\rangle_R |E_n\rangle_L =0$), their expectation values are finite, hence $H_0$ is well defined.

 But the fact that    $H'_L$ and $H'_R$ are not well defined  indicates that a factorization of the Hilbert space into the Hilbert spaces defined on $R$ and $L$ is not well defined in the $G\to 0$ limit.   
  We can compare our results, $\langle \Phi| H_R |\Phi\rangle\sim {\cal O}(G^{-1})$ and $\delta H_R\sim {\cal O}(G^{-1/2})$, with the boundary Hamiltonian of the eternal AdS black hole, which describes  the ${\cal N}=4$ super Yang-Mills theory \cite{Witten:2021unn}. 
  In the large $N$ limit, Hamiltonians in the left and right boundaries $H_L$ and $H_R$ have thermal expectation values of ${\cal O}(N^2) \sim {\cal O}(G^{-1})$ and  $H'_{L/R}=H_{L/R}-\langle H_{L/R}\rangle$ satisfy $\langle {H'_{L/R}}^2\rangle \sim{\cal O}(N^2)$ hence  $\delta H_{L/R} \sim {\cal O}(N)\sim {\cal O}(G^{-1/2})$, showing the same behaviors as $\langle \Phi| H_R |\Phi\rangle$ and $\delta H_R$, respectively.



\section{von Neumann algebra for inflationary quasi-dS space }
\label{sec:II1quasidS}

\subsection{von Neumann  algebra for quasi-dS space}
\label{sec:II1dS}

 In order to find the   von Neumann algebra description of quasi-dS space, we first consider  Type II$_1$ algebra for dS space discussed in \cite{Chandrasekaran:2022cip} and see how it is modified by the nonzero energy flux across the horizon we obtained in Section \ref{sec:hordyn}.
 Since a static observer can access the static patch $R$ only, the quantum description of (quasi-)dS space as seen by the static observer is made in terms of the local observables on $R$.
  Moreover, in the dS limit,  the static patch is invariant under the subgroup of the dS isometry consists of the static time translation and rotation hence operators on the static patch are required to be invariant under the subgroup.
  However, as pointed out in \cite{Chandrasekaran:2022cip}, the only operators that commute with the static time translation generator are those proportional to the identity.
In order to resolve this issue,  \cite{Chandrasekaran:2022cip}   suggested that the nontrivial operators can be considered by taking the Hamiltonian of the static observer into account in addition, such that the  total Hamiltonian is given by $H_0+\widehat{q}$, where $H_0$ is the static time translation generator  and $\widehat{q}$ is the observer Hamiltonian. 
To see the meaning of $\widehat{q}$, we note that the observer detects the thermal radiation using, for example, the apparatus consisting of the large number of atoms.
  In this case, the energy eigenvalues $q$ are almost continuous, forming the band structure such that we can detect the thermal radiation with any frequency  by observing  the transition between the  energy levels of the apparatus.
  Since the energy levels of the apparatus described above is also bounded from below, it is reasonable to assume that the eigenvalues $q$ are nonnegative and $\widehat{q}$  acts on $L^2(\mathbb{R}^+)$, the Hilbert space of the square integrable function of  $q$.
  Moreover, we can define the time measured by the observer's clock along the   observer's worldline, which corresponds to the eigenvalue of the operator $\widehat{q}$ conjugate to $\widehat{p}$. 

 For quasi-dS space, by taking the inflationary period to be $(\epsilon_H H)^{-1}$, the nonzero energy flux  across the horizon $\langle \Phi | H_R|\Phi\rangle$ becomes divergent in the $G\to 0$ limit.
 As discussed in Section \ref{sec:hordyn},  $H'_R=H_R-\langle \Phi | H_R|\Phi\rangle$, the renormalized static time translation generator (restricted to the static patch $R$)    is not well defined since $\langle \Phi | (H'_R)^2|\Phi\rangle \sim {\cal O}(G^{-1})$ also diverges  in the $G\to 0$ limit.
 The similar problem also arises in the boundary Hamiltonian of the AdS black hole, in which the issue is circumvented by considering $H'_R/N$ since ${\cal O}(N)$ is equivalent to ${\cal O}(G^{-1/2})$ in the large $N$ limit \cite{Witten:2021unn}.
Motivated by this, we may define $H'_R/N$ where $N$ is now the dimensionless parameter of ${\cal O}(G^{-1/2})$, say, $M_{\rm Pl}/H$.
Then following \cite{Witten:2021unn}, $H'_R/N$ can be expressed as
\dis{\frac{H'_R}{N}=U + \frac{H_0}{N},}
where $U$ is an operator which commutes with any observables on $R$.
Whereas $U$ is in fact $H'_L/N$, it is also identified with $H'_R/N$ in the $G\to 0$ limit, in which $H_0/N \to 0$ and $[a, H'_R/N]=(i/N)\partial a/\partial t_s \to 0$ for any operator $a$ on $R$.
Moreover, the divergence of $\langle \Phi | H_R|\Phi\rangle/N \sim {\cal O}(G^{-1/2})$ in the $G\to 0$ limit indicates that the lower bound on $U$ is $-\infty$ thus the eigenvalue of $U$ can take  any real number  in $(-\infty, +\infty)$ and the Hilbert space relevant to  $U$ is given by $L^2(\mathbb{R})$.
In the following discussion, we will take the factor $1/N$ to be implicit for convenience. 
Indeed, as pointed out in \cite{Witten:2021unn}, if we are interested in the ordinary  functions of $N$, we may work with $H'_R$ instead of $H'_R/N$. 
\footnote{Our discussion is based on the canonical ensemble in which the fluctuation $\delta H_R\sim {\cal O}(G^{-1/2})$ is divergent in the $G\to 0$ limit.
On the other hand, we don't need to divide $H_R$ by $N$ in the microcanonical ensemble as the fluctuation in $H_R$ is restricted to be ${\cal O}(1)$ \cite{Chandrasekaran:2022eqq}.
The reason we do not consider the microcanonical ensemble is that the divergent fluctuation in $H_R$ is induced by the fluctuation in the curvature perturbation and so far as we know there is no physical reason to restrict the fluctuation to be ${\cal O}(1)$. 
}
Defining $X\equiv NU=H'_L$, $H'_R$ is written as $X+ H_0$ with $X$ acting on the Hilbert space $L^2(\mathbb{R})$ and   the (renormalized) Hamiltonian restricted to the static patch $R$ is given by $\widehat{H}=H_0+X+\widehat{q}$.

 Now we can construct Type III algebra  ${\cal A}_R \otimes  B(\mathbb{R}) \otimes  B(\mathbb{R}^+)$, where ${\cal A}_R$ is the algebra of the observables on $R$ and $B(\mathbb{R}^{(+)})$ is the algebra of bounded operators acting on $\mathbb{R}^{(+)}$.
 This is converted into Type II algebra by imposing the diffeomorphism invariance as a gauge constraint.
Focusing on an (approximate) isometry of the static time translation, the algebra of the  observables on $R$ is given by an  invariant subalgebra with respect to $\widehat{H}$, 
 \dis{\widehat{\cal A}_R=\big({\cal A}_R\otimes  B(\mathbb{R}) \otimes  B(\mathbb{R}^+)\big)^{\widehat H}. }  
 The elements of  $\widehat{{\cal A}}_R$ can be explicitly written by introducing an operator $\widehat{p}$ conjugate to $\widehat{q}$ satisfying $[\widehat{q}, \widehat{p}]=i$, which is interpreted as a (fluctuating) time   measured by the static observer.
By requiring $[\widehat{q}, H_0]=0$, $\widehat{q}$ belongs to $\widehat{{\cal A}}_R$. 
Moreover, for any $a\in {\cal A}_R$, one finds that its gravitational dressing or  the outer automorphism, 
\dis{{a}'=e^{i (H_0+X) \widehat{p}} a e^{-i (H_0+X) \widehat{p}}\label{eq:a'}}
satisfies $[\widehat{q}, {a}']=-[H_0+X, {a}']$, or equivalently, $[\widehat{H}, {a}']=0$.
Therefore, $\widehat{{\cal A}}_R$  is generated by
 \dis{\{ {a}'=e^{i (H_0+X) \widehat{p}} a e^{-i (H_0+X) \widehat{p}}, \widehat{q}\}. }
 By taking the conjugation by $e^{-i (H_0+X) \widehat{p}}$, one finds that it is equivalent to $\{a, \widehat{q}-(H_0+X)\}$.

 In order to implement the finite (renormaliazed) entropy, we need to define `trace' in a sensible way.
  The trace here refers to, in a somewhat abstract sense,  a linear functional of operators satisfying ${\rm Tr}(ab)={\rm Tr}(ba)$ and ${\rm Tr}(a^\dagger a)>0$ for a nonzero $a$.
  This can be used to defined the `renormalized' (thus finite) entropy in dS space (for further discussion, see \cite{Witten:2018zxz} and \cite{Chandrasekaran:2022cip}, which is also reviewed in the paragraph containing \eqref{eq:TrdS}). 
  Just like the case of perfect dS space, the trace can be defined in terms of the Bunch-Davies state $|\Psi\rangle$ which is invariant under the dS isometries.  
\footnote{  We note that since $H_0$ belongs to the isometry generators, $H_0|\Psi\rangle=0$ is satisfied and   $|\Psi\rangle$ does not distinguish $\widehat{q}-H_0$ from $\widehat{q}$.
 From this and $[X, a]=0$, one finds that for $a'$ defined in \eqref{eq:a'}, $\langle \Psi|a'|\Psi\rangle$ is identified with $\langle \Psi|a|\Psi\rangle$.  }
 Then we  define the trace of  any operator $\widehat{a} \in \widehat{\cal A}_R$  by  
  \dis{{\rm Tr}(\widehat{a})=  \int_{-\infty}^{\infty}\beta dx e^{\beta x}\int_0^\infty \beta dq  e^{-\beta q }\langle\Psi|\widehat{a}|\Psi\rangle,\label{eq:Trace}} 
  where $x$ is the eigenvalue of $X$. 
  
For perfect dS space, since the divergent fluctuation in $X$ is not taken into account,  the integration over $x$ is absent and  the trace of the identity  is not divergent but finite :  ${\rm Tr}(1)=1$.
To see the physical meaning of the identity in this case, let us observe the trace of other operators in $\widehat{{\cal A}}_R$ generated by $\{a, \widehat{q}-H_0 \}$. 
For $a\in {\cal A}_R$ which is independent of $\widehat{q}-H_0$,
 \dis{&{\rm Tr}(a)={\rm Tr}(a 1)=\Big(\beta\int_0^\infty dq e^{-\beta q}\Big)\langle\Psi|a|\Psi\rangle = \langle\Psi|a|\Psi\rangle,\label{eq:TrdS}}
 showing that ${\rm Tr}(a)$ is the expectation value of the   local operator $a$ with respect to $|\Psi\rangle$.
 On the other hand, when the operator $G\in \widehat{{\cal A}}_R$ is given by a function of $\widehat{q}-H_0$,   
  \dis{{\rm Tr}(G(\widehat{q}-H_0))&={\rm Tr}(G(\widehat{q}-H_0) 1)= \beta\int_0^\infty dq e^{-\beta q} \langle\Psi| G(\widehat{q}-H_0)|\Psi\rangle
 \\
 & = \beta\int_0^\infty dq e^{-\beta q} \langle\Psi| G(q)|\Psi\rangle = \beta\int_0^\infty dq e^{-\beta q}  G(q),}
 from which one finds that for Tr$(G)$ to be an expectation value, $\beta e^{-\beta q}$ is interpreted as the probability distribution of the eigenvalues of $\widehat{q}$. 
 Then it is reasonable to interpret the identity as a density matrix describing the maximal entanglement :  the entropy of the system in the perfect dS background is maximized at  $-{\rm Tr}(1 \log 1)=0$. 
  In other words, among the states in the Hilbert space ${\cal H}\otimes L^2(\mathbb{R}^+)$ on which the algebra $\widehat{{\cal A}}_R$ acts,
 \dis{|\Psi_{\rm max}\rangle =|\Psi\rangle \otimes \int_0^\infty dq \sqrt{\beta}e^{-\beta {q}/2}|q\rangle} 
 gives the maximal entropy as the density matrix $\rho_{\rm max}=1$ is obtained from  $\langle \Psi_{\rm max}|\widehat{a}|\Psi_{\rm max}\rangle={\rm Tr}(\rho_{\rm max}\widehat{a})$ \cite{Chandrasekaran:2022cip}.
 This is a feature of Type II$_1$ von Neumann algebra. 

In contrast, for  quasi-dS space, the integration over $x$ in \eqref{eq:Trace} leads to Tr$(1)=\infty$, which means that the density matrix for the maximal entanglement is not renormalized.
Then $\widehat{{\cal A}}_R$ belongs to Type II$_\infty$ von Neumann algebra, in which the trace is sensibly defined  only for the subset of the algebra.


\subsection{Density matrix and entropy}
\label{sec:rhoandS}

Since the inflationary quasi-dS background slightly breaks the dS isometry, the quantum state during inflation  $|\Phi\rangle$ is no longer the Bunch-Davies state $|\Psi\rangle$.  
In order to find the density matrix in this case, we   consider a state
\dis{&|\widehat{\Phi}\rangle=|\Phi\rangle \otimes g(X)\otimes f(\widehat{q})}
 in ${\cal H}\otimes L^2(\mathbb{R})\otimes L^2(\mathbb{R}^+)$, where
\dis{&g(X)=\int_{-\infty}^\infty g(x)|x\rangle,\quad\quad
 f(\widehat{q})=\int_{0}^\infty f(q)|q\rangle.}
Here   $|g(x)|^2$  and $|f(q)|^2$ are interpreted as  the probability distributions of   $x$  and $p$  with the normalizations $\int_{-\infty}^\infty |g(x)|^2dx=1$ and $\int_{0}^\infty |f(q)|^2dq=1$, which are assumed to be  slowly varying functions of $x$ and $q$, respectively. 
Then $\rho_{\widehat{\Phi}}$, the density matrix associated with $|\widehat{\Phi}\rangle$ is defined as
\dis{\langle \widehat{\Phi} | \widehat{a} |\widehat{\Phi}\rangle={\rm Tr}(\rho_{\widehat{\Phi}}\widehat{a})=\int_{-\infty}^{\infty}\beta dx e^{\beta x} \langle \Psi_{\rm max} | \rho_{\widehat{\Phi}}\widehat{a} |\Psi_{\rm max}\rangle, \label{eq:defrhoPhi}}
for any $\widehat{a}\in \widehat{\cal A}_R$.

 In order to obtain $\rho_{\widehat{\Phi}}$, we need to convert the states written in terms of $|\Psi\rangle$, i.e.,   the states constructed by acting the operators in ${\widehat{\cal A}_R}$ on $|\Psi\rangle$, into those written in terms of $|\Phi\rangle$.
 This is  well described by the relative Tomita operator $S_{\Phi|\Psi}$, an antilinear operator satisfying
\footnote{While we follow the notations in \cite{Chandrasekaran:2022cip}, they are different from those in \cite{Witten:2018zxz} : $S_{\Phi|\Psi}$ and $\Delta_{\Phi|\Psi}$ in \cite{Chandrasekaran:2022cip} are $S_{\Psi|\Phi}$ and $\Delta_{\Psi|\Phi}$ in \cite{Witten:2018zxz}, respectively.}
 \dis{S_{\Phi|\Psi} a |\Psi\rangle= a^\dagger |\Phi\rangle}
 for all $a\in {\cal A}_R$.
 From this, we define the relative modular operator
 \dis{\Delta_{\Phi|\Psi} = e^{-h_{\Phi|\Psi}} = S_{\Phi|\Psi}^\dagger S_{\Phi|\Psi},}
 which gives the relation
 \dis{\langle\Psi|\Delta_{\Psi|\Phi}a|\Psi\rangle 
 &= \langle\Psi|S_{\Phi|\Psi}^\dagger S_{\Phi|\Psi} a|\Psi\rangle = \langle\Psi|S_{\Phi|\Psi}^\dagger  a^\dagger|\Phi\rangle = \langle \Phi|a^\dagger |\Phi\rangle^*
 \\
 &=\langle \Phi|a|\Phi\rangle. \label{eq:relmod}}
In the same way, we can also define the Tomita operators and the modular operators  $\Delta_\Phi =e^{-h_\Phi}=S_\Phi^\dagger S_\Phi$ and $\Delta_\Psi =e^{-h_\Psi}=S_\Psi^\dagger S_\Psi$ satisfying
 \dis{S_\Phi a|\Phi\rangle=a^\dagger |\Phi\rangle,\quad
 S_\Psi a|\Psi\rangle=a^\dagger |\Psi\rangle,}
 respectively. 
 In order to find the physical meaning of these modular operators, in addition to ${\cal A}_R$, the local algebra restricted to the static patch $R$, we consider another local algebra ${\cal A}'_R$, a commutant of ${\cal A}_R$. 
 When we  extend the spacetime manifold to  $L$ and $B$, ${\cal A}'_R$ can be given by the local algebra on the complementary static patch $L$.
 Then it was shown that (see, e.g., Section IV. A of \cite{Witten:2018zxz}) the density matrix $\rho_\Psi$ for algebra ${\cal A}_R$ and $\rho'_\Psi$ for ${\cal A}'_R$ associated with the state $|\Psi\rangle$ satisfy
 \dis{\Delta_\Psi =\rho_\Psi \otimes {\rho'_\Psi}^{-1}.}
 We can also find the similar relations $\Delta_\Phi =\rho_\Phi \otimes {\rho'_\Phi}^{-1}$, $\Delta_{\Psi|\Phi}=\rho_\Psi\otimes {\rho'_\Phi}^{-1}$, and $\Delta_{\Phi|\Psi}=\rho_\Phi\otimes {\rho'_\Psi}^{-1}$.
  Given the static time translation generators $H_R$ for ${\cal A}_R$ and $H_L$ for ${\cal A}'_R={\cal A}_L$, the density matrices can be written as
\dis{\log\rho_\Psi=-\beta H_R +C,\quad \log\rho'_\Psi=-\beta H_L +C,\label{eq:rhorho'}}   
 respectively, from which one finds that $h_\Psi=\beta (H_R-H_L)$. 
  Since the static time in $R$ runs in the opposite direction  to  that in $L$, $H_R-H_L$ is nothing more than the total static time translation generator $H_0$.
  Thus, $h_\Psi$ is identified with $\beta H_0$.
The explicit value of $C$ in \eqref{eq:rhorho'} can be obtained by taking the expectation value with respect to $|\Psi\rangle$.
 From $\langle \Psi|H_R|\Psi\rangle=0$ (no energy flux across the horizon) and $\langle\Psi|\log\rho_\Psi|\Psi\rangle={\rm Tr}(\rho_\Psi\log\rho_\Psi)=-S_{\rm bulk}(R)_\Psi$ where $R$ denotes  the bulk of the static patch $R$, we obtain
 \dis{\log \rho_\Psi=-\beta H_R-S_{\rm bulk}(R)_\Psi.\label{eq:rhopsi}}

 Comparing  \eqref{eq:defrhoPhi} and \eqref{eq:relmod}, it is reasonable to expect that $\rho_{\widehat{\Phi}}$ contains  $\Delta_{\Phi|\Psi}$, which converts the expectation value with respect to $|\Psi\rangle$ into that with respect to $|\Phi\rangle$.
 This indeed is supported by the relation $\Delta_{\Phi|\Psi}=\rho_\Phi\otimes {\rho'_\Psi}^{-1}$ and the observation that $L\cup B$ is a  copy of $R\cup T$ thus $\rho'_\Psi$ is just given by $\rho_\Psi$.
 The rest part of $\rho_{\widehat{\Phi}}$ depends on the probability distribution of $x$ and   $q$.
 Moreover, the fact that the observer's energy is described in a probabilistic way implies that the observer's time has an uncertainty.
 In \cite{Chandrasekaran:2022cip} and \cite{Chandrasekaran:2022eqq}, it was argued that when the fluctuation in the observer's time is bounded by $\varepsilon \ll \beta$ and $g(x)$ as well as $f(q)$ is  slowly varying  over $1/\varepsilon$, i.e., $|\Delta x|, |\Delta q| \sim 1/\varepsilon$, the density matrix $\rho_{\widehat{\Phi}}$ is given by
 \dis{\rho_{\widehat{\Phi}} = \frac{1}{\beta^2}\big| g(X+h_\Psi/\beta) f ( \widehat{q})\big|^2e^{\beta (-X+\widehat{q})}\Delta_{\Phi|\Psi}+{\cal O}(\varepsilon).\label{eq:rhoPhi}}
The justification of \eqref{eq:rhoPhi} can be found in Appendix \ref{app:derden}.
We note that in quasi-dS space, the curvature perturbation induces  the fluctuation in $\phi(t)$ hence that in the flat time given by $\delta t=\delta \phi/\dot{\phi}(t)$. 
As can be inferred from \eqref{eq:flucphi}, these fluctuations are accumulated as time goes on, hence  negligibly small compared to the fluctuation in the static observer's time provided
\dis{\delta t=\frac{\delta \phi}{\dot{\phi}(t)}=\sqrt{\frac{1}{\pi\epsilon_H}}\frac{(H\Delta t)^{1/2}}{M_{\rm Pl}} < \varepsilon \ll \beta=\frac{2\pi}{H}.}
If the inequality is satisfied until $\Delta t \sim \beta$, i.e., several $e$-folds, $\epsilon_H$ is required to be larger than $H^2/M_{\rm Pl}^2$, which is known as the condition that the eternal inflation does not take place.
That is, if $\epsilon_H$ is too small, the field value of $\phi(t)$ strongly fluctuates and $H$ may stay at some constant value for a long time instead of decreasing in time through the slow-roll.  
In this case, the semiclassical description we have considered is no longer valid.
If the inequality is satisfied until the end of inflation $\Delta t \sim (\epsilon_H H)^{-1}$, we have the stronger bound $\epsilon_H > H/M_{\rm Pl}$.

Then the von Neumann entropy of the static patch    associated with the state $|\widehat{\Phi}\rangle$, which will be identified with the generalized entropy up to the addition of constant, is written as
\dis{S(R)_{\widehat{\Phi}}&=-\langle\widehat{\Phi}|\log\rho_{\widehat{\Phi}}|\widehat{\Phi}\rangle
\\
&=\langle \widehat{\Phi}|h_{\Phi|\Psi}|\widehat{\Phi}\rangle- \langle\widehat{\Phi}|(\beta(\widehat{q}-X) +h_\Phi)|\widehat{\Phi}\rangle
\\
&\quad
-\int_{0}^\infty dq |f(q)|^2(\log|f(q)|^2-\log\beta)
-\int_{-\infty}^\infty dx |g(H_R)|^2(\log|g(H_R)|^2-\log\beta),}
where in the second term $\langle\Phi|h_\Phi|\Phi\rangle=0$ which is obtained from $S_\Phi|\Phi\rangle=|\Phi\rangle$ is added and $H_R$ indicates $X+h_\Psi/\beta=X+H_0$. 
From   $\Delta_\Psi=\rho_\Psi \otimes {\rho'_\Psi}^{-1}$ and $\Delta_{\Psi|\Phi}=\rho_\Psi\otimes {\rho'_\Phi}^{-1}$, one finds  $\Delta_{\Phi|\Psi}^{is}\Delta_\Psi^{-is}=\Delta_\Phi^{is}\Delta_{\Psi|\Phi}^{-is}$ (this quantity is called the Connes cocycle), which leads to $h_{\Phi|\Psi}-h_\Phi=h_\Psi-h_{\Psi|\Phi}$.
Then $S (R)_{\widehat{\Phi}}$ is rewritten as
\dis{S (R)_{\widehat{\Phi}}=&-\langle \widehat{\Phi}|h_{\Psi|\Phi}|\widehat{\Phi}\rangle- \langle\widehat{\Phi}|(\beta(\widehat{q}-X)-h_\Psi)|\widehat{\Phi}\rangle
\\
&-\int_{0}^\infty dq |f(q)|^2(\log|f(q)|^2-\log\beta)
-\int_{-\infty}^\infty dx |g(H_R)|^2(\log|g(H_R)|^2-\log\beta).\label{eq:SRPhi}}
Let us first consider the second line.
The first integral is interpreted as the entropy of the static observer, which is evident from the fact that $|f(q)|^2$ is the probability distribution of $\widehat{q}$ eigenvalues, the observer's energy.
As for the last term, we recall that the fluctuation in $H_R$ originates from the fluctuation in $\phi(t)$, or equivalently, the flat time $t$.
Since $H$ also varies depending on $t$, the value of  $H$   at the end of inflation also fluctuates, the probability distribution of which is described by $|g(H_R)|^2$. 
These two integrals in the second line are not explicitly relevant to the excitations of matter in the static patch, and can be identified with $S_{\rm bulk}(\infty)_{\widehat{\Phi}}$, the bulk entropy associated with $|\widehat{\Phi}\rangle$ at the end of inflation.
This is because at late time, the wavelength  of almost all the excitations will be stretched beyond the horizon so the static observer does not find any excitation except for that of the observer state inside the horizon. 
We can compare it with the static observer in the Bunch-Davies state $|\Psi\rangle$.
As $|\Psi\rangle$ is invariant under the static time translation generated by $H_0$, the bulk entropy will be constant in time, i.e.,  $S_{\rm bulk}(R)_{\widehat{\Psi}}=S_{\rm bulk}(\infty)_{\widehat{\Psi}}$, and it measures the entropy of empty dS space without any excitation except for that of the observer state.
 For the state during inflation $|\Phi\rangle$, in contrast, the background does not respect the isometry generated by $H_0$ any longer.
Then the spontaneous breaking of the isometry by the background gives rise to the curvature perturbation which does not appear in perfect dS space. 
 But the background is still close to dS space and the wavelength of these excitations would be stretched as the universe undergoes   accelerated expansion.
 Then just like the perfect dS background, almost all the excitations cross the horizon after several $e$-folds  and  only the fluctuations of  the observer's energy and  the value of $H$ contribute to the entropy.
 Then $S_{\rm bulk}(\infty)_{\widehat{\Phi}}$ can be identified with the sum of $S_{\rm bulk}(\infty)_{\widehat{\Psi}}=S_{\rm bulk}(R)_{\widehat{\Psi}}$ and the last integral in \eqref{eq:SRPhi}.
 The same argument leads to $S_{\rm bulk}(\infty)_{{\Phi}}\simeq S_{\rm bulk}(\infty)_{{\Psi}}=S_{\rm bulk}(R)_{{\Psi}}$.

 We also note that the last two integrals in \eqref{eq:SRPhi} reflects two different ways to give rise to the uncertainty of the horizon area.
 First, since the horizon is deformed by the backreaction of observer's energy $q$, the   horizon area fluctuates as    $q$ fluctuates   \cite{Susskind:2021omt, Susskind:2021dfc, Chandrasekaran:2022cip}.
 Second, as we remarked earlier, the probability distribution  $|g(H_R)|^2$ is induced by the  fluctuation in time, which leads to the fluctuation in the horizon radius, thus that in the horizon area at late time.

 Summarizing the discussion so far, $S (R)_{\widehat{\Phi}}$ can be written as
 \dis{S (R)_{\widehat{\Phi}}=-\langle \widehat{\Phi}|h_{\Psi|\Phi}|\widehat{\Phi}\rangle- \langle\widehat{\Phi}|(\beta(\widehat{q}-X)-h_\Psi)|\widehat{\Phi}\rangle + S_{\rm bulk}(\infty)_{\widehat{\Phi}}.\label{eq:Sbulkbinf}}
 As we will see in the next section, whereas the first term is identified with the negative of the change in the generalized entropy, the term $- \langle\widehat{\Phi}|(\beta(\widehat{q}-X)-h_\Psi)|\widehat{\Phi}\rangle = \langle\widehat{\Phi}|(H_R-\beta\widehat{q})|\widehat{\Phi}\rangle$ can be interpreted as the deformation of the horizon area from the initial value.
 This leads to the relation $S (R)_{\widehat{\Phi}}=S_{\rm gen}(R)_{\widehat{\Phi}}+$(constant).

\subsection{Horizon dynamics and entropy in quasi-dS space}
\label{sec:extquasidS}

 We now focus on the first term of the RHS in \eqref{eq:Sbulkbinf}.
 It was argued in \cite{Chandrasekaran:2022eqq} that this term is identified with a negative of the relative entropy, $-S_{\rm rel}(\Phi || \Psi)$.
 Indeed, this is the case if we consider the expectation value of  $-h_{\Psi|\Phi}$ with respect to $|\Phi\rangle$,
\dis{-\langle \Phi | h_{\Psi|\Phi}|\Phi\rangle&= \langle \Phi | \log(\Delta_{\Psi|\Phi})|\Phi\rangle
=\langle \Phi | (\log(\rho_\Psi)-\log(\rho_\Phi)) |\Phi\rangle
\\
&={\rm Tr}(\rho_\Phi\log\rho_\Psi)-{\rm Tr}(\rho_\Phi\log\rho_\Phi)=-S_{\rm rel}(\Phi||\Psi).}
Moreover, the relative entropy can be rewritten as 
\cite{Wall:2010cj, Chandrasekaran:2022eqq}
 \dis{S_{\rm rel}( {\Phi} || {\Psi})=S_{\rm gen}(\infty)_{ {\Phi}}-S_{\rm gen}(R)_{ {\Phi}}\label{eq:Srelrel},}
 i.e., the difference  between the generalized entropies at initial ($U=0)$ and  late ($U=\infty)$ times given by 
 \dis{S_{\rm gen}(R)=\frac{A(b)}{4G}+S_{\rm bulk}(R),\quad
 S_{\rm gen}(\infty)=\frac{A(b_\infty)}{4G}+S_{\rm bulk}(\infty),}
 respectively, where $b$ and $b_\infty$ indicate  the horizon cuts  at initial and late times  as depicted in Figure \ref{Fig:dSpenrose}.
 This comes from the observation that $H_R$, the static time translation generator restricted to the static patch $R$  (recall that $H_0=h_\Psi/\beta=H_R-H_L$) satisfies
 \dis{\langle  {\Phi} | \beta H_R| {\Phi}\rangle 
 =\frac{A(b_\infty)}{4G}-\frac{A( b)}{4G}.\label{eq:areadiff}}
 Then from $S_{\rm bulk}(\infty)_\Phi \simeq S_{\rm bulk}(\infty)_\Psi=S_{\rm bulk}(R)_\Psi$ we obtain
 \dis{S_{\rm gen}(\infty)_\Phi-S_{\rm gen}(R)_\Phi
 &=\frac{A(b_\infty)}{4G}-\frac{A( b)}{4G}+S_{\rm bulk}(\infty)_\Phi-S_{\rm bulk}(R)_\Phi
 \\
 &\simeq \frac{A(b_\infty)}{4G}-\frac{A( b)}{4G}+S_{\rm bulk}(R)_\Psi-S_{\rm bulk}(R)_\Phi.}
 From \eqref{eq:rhopsi}, the change in the horizon area can be written as
 \dis{\frac{A(b_\infty)}{4G}-\frac{A( b)}{4G}&=\beta\langle \Phi|H_R|\Phi\rangle=-\langle \Phi|\log\rho_\Psi|\Phi\rangle-S_{\rm bulk}(R)_\Psi, \label{eq:Areachange}}
from which the above expression is rewritten as
 \dis{S_{\rm gen}(\infty)_\Phi-S_{\rm gen}(R)_\Phi &=
 -\langle \Phi|\log\rho_\Psi|\Phi\rangle-S_{\rm bulk}(R)_\Phi
 \\
 &=-\langle \Phi|\log\rho_\Psi|\Phi\rangle+\langle \Phi|\log\rho_\Phi|\Phi\rangle =S_{\rm rel}(\Phi||\Psi),}
 which confirms \eqref{eq:Srelrel}.

Meanwhile, when we replace $|\Phi\rangle$ by $|\widehat{\Phi}\rangle$, the first term  in \eqref{eq:Sbulkbinf} is given by
\dis{-\langle \widehat{\Phi} | h_{\Psi|\Phi}|\widehat{\Phi}\rangle&=
\langle \widehat{\Phi} | (\log \rho_\Psi -\log \rho_\Phi ) |\widehat{\Phi}\rangle
\\
&=-\langle \widehat{\Phi} | \beta H_R|\widehat{\Phi}\rangle -S_{\rm bulk}(R)_\Psi  -\langle \widehat{\Phi} | \log\rho_\Phi|\widehat{\Phi}\rangle,}
where again  the expression \eqref{eq:rhopsi} for $\log \rho_\Psi$ is used  for the first two terms in the second line.
To proceed, we note that the relation \eqref{eq:areadiff} holds even if  $|\Phi\rangle$ is replaced by $|\widehat{\Phi}\rangle$, with the explicit values of $A(b_\infty)$ and $A(b)$ are changed reflecting the backreaction of the observer.
Moreover, since $\log\rho_\Phi$ contains the probability distribution of the matter excitations in the bulk,  it is tempting to relate the   last term $ -\langle \widehat{\Phi} | \log\rho_\Phi|\widehat{\Phi}\rangle = -{\rm Tr}(\rho_{\widehat{\Phi}}\log\rho_\Phi)$ to the bulk entropy. 
However, it cannot be identified with $S_{\rm bulk}(R)_{\widehat{\Phi}}$ as the dynamics of the bulk   in the state $|\widehat{\Phi}\rangle$ is also affected by the fluctuations in $q$ and $x$, which are not reflected in $\log\rho_\Phi$.
Since these fluctuations remain until the end of inflation when all the matter excitations cross the horizon, if we conjecture that the difference $S_{\rm bulk}(R)_{\widehat{\Phi}}-(-{\rm Tr}(\rho_{\widehat{\Phi}}\log\rho_\Phi))$ which contains the  effects of the fluctuations in $q$ and $x$ on the bulk dynamics  are time independent at leading order, it can be identified with  $S_{\rm bulk}(\infty)_{\widehat{\Phi}}-S_{\rm bulk}(\infty)_{{\Phi}}$.
Then we obtain
\dis{-\langle \widehat{\Phi} | h_{\Psi|\Phi}|\widehat{\Phi}\rangle &=\Big( -\frac{A(b_\infty)}{4G}+\frac{A( b)}{4G}\Big)-S_{\rm bulk}(R)_\Psi + \big(S_{\rm bulk}(R)_{\widehat{\Phi}}-S_{\rm bulk}(\infty)_{\widehat{\Phi}}+S_{\rm bulk}(\infty)_{{\Phi}}\big)
\\
&=-\Big(\frac{A(b_\infty)}{4G}+S_{\rm bulk}(\infty)_{\widehat{\Phi}}\Big)+\Big(\frac{A( b)}{4G}+S_{\rm bulk}(R)_{\widehat{\Phi}}\Big)
\\
&=-\big(S_{\rm gen}(\infty)_{\widehat{\Phi}}-S_{\rm gen}(R)_{ \widehat{\Phi}}\big),}
where in the second line we use the relation $S_{\rm bulk}(\infty)_{{\Phi}} \simeq S_{\rm bulk}(\infty)_{{\Psi}}=S_{\rm bulk}(R)_{{\Psi}}$.

 Now we investigate the change in the horizon area more explicitly.
  When   the observer's energy  $q$ in perfect dS space is  concentrated in the tiny region, the Schwarzschild-de Sitter black hole can be created and the metric is modified as  
 \footnote{There has been a conjecture that in the absence of an observer collecting information, quantum gravity forbids the production of the black hole through the fluctuation \cite{Cohen:1998zx}.
 For discussions on how this conjecture applies to the dS background, see, e,g, \cite{Banks:2019arz, Seo:2021bpb, Castellano:2021mmx, Seo:2022uaz}. }
 \dis{&ds^2=-f(r_s)dt_s^2+\frac{1}{f(r_s)}dr_s^2+r_s^2(d\theta^2+\sin^2\theta d\phi^2),
 \\
 &f(r_s)=1-\frac{2G q}{r_s}-H^2 r_s^2.}
 If $q$ is small enough, say, $q\lesssim H$ or $q/\Delta E \sim H^2/M_{\rm Pl}^2 \ll 1$, the linear expansion in $q$ is valid such that the   cosmological horizon and the black hole radii are   approximated as $H^{-1}-Gq$  and $2Gq$, respectively.
 For quasi-dS space, the radius of the cosmological horizon   deformed  by the slow-roll as well as the backreaction of the observer's energy is given by $r_H=H^{-1}+\epsilon_H\Delta t  -Gq$.
 Here $H$   is the initial value of the Hubble parameter,   and the Hubble parameter during the inflationary period can be approximated by this constant value.
 Then  the  (cosmological) horizon area at initial time is estimated as
 \dis{\frac{A(  b)}{4G} \simeq  \frac{\pi}{G}(H^{-1}-Gq)^2 \simeq 
 \frac{\pi}{GH^2} - \beta q,\label{eq:areaobs}}
and that at late time, namely, just after the inflationary period $\Delta t\simeq (\epsilon_H H)^{-1}$,  is approximated as
\dis{\frac{A(b_\infty)}{4G} \simeq \frac{\pi}{G}(H_f^{-1}-Gq)^2 \simeq 
 \frac{\pi}{GH_f^2} - \beta_f q,}
where $H_f^{-1}=H^{-1} +\epsilon_H\Delta t$ is the value of the Hubble radius at the end of inflation and $\beta_f=(2\pi)/H_f$.
 As we have seen in Section \ref{sec:hordyn}, the change in the horizon area leads to the energy flux across the horizon as $\langle \widehat{\Phi}| H_R |\widehat{\Phi}\rangle=\int dt \beta^{-1}(d/dt)[A/(4G)]$, or symbolically, $\langle \widehat{\Phi}| \beta H_R |\widehat{\Phi}\rangle =[A(b_\infty)-A(b)]/4G$.
 Since $X+H_0=X+h_\Psi/\beta = H_R$,  \eqref{eq:Sbulkbinf} is rewritten as
  \dis{S (R)_{\widehat{\Phi}}&=-(S_{\rm gen}(\infty)_{\widehat{\Phi}}-S_{\rm gen}(b)_{\widehat{\Phi}})- \langle\widehat{\Phi}|\beta\widehat{q} |\widehat{\Phi}\rangle +\Big(\frac{A(b_\infty)}{4G}-\frac{A(b )}{4G}\Big)+ S_{\rm bulk}(\infty)_{\widehat{\Phi}}
  \\
  &=S_{\rm gen}(b)_{\widehat{\Phi}}- \langle\widehat{\Phi}|\beta\widehat{q} |\widehat{\Phi}\rangle-\frac{A(b )}{4G}. }
  We also note that the state $|\widehat{\Phi}\rangle$ contains the probability distribution of the observer's energy $q$, in which  $-\beta q$ in \eqref{eq:areaobs} is modified to $- \langle\widehat{\Phi}|\beta\widehat{q} |\widehat{\Phi}\rangle$.
  Therefore, we arrive at
  \dis{S (R)_{\widehat{\Phi}} = S_{\rm gen}(b)_{\widehat{\Phi}}-\frac{\pi}{GH^2},}
 and since $H$ is a constant, $S (R)_{\widehat{\Phi}}$ is identified with $S_{\rm gen}(b)_{\widehat{\Phi}}$ up to the addition of a constant.

 \section{Conclusion}
\label{sec:conclusion}

 Throughout this article, we have investigated how Type II$_1$ von Neumann algebra description of perfect dS space is modified in the inflationary quasi-dS space.
 Unlike perfect dS space, quasi-dS space allows the nonvanishing energy flux across the horizon, which is identified with the expectation value of the static time translation generator.
 In the evaluation of the energy flux, we assume  the inflationary period to be $(\epsilon_H H)^{-1}$, which is natural in the sense that after this time scale $H$ deviates significantly from the initial value hence it is no longer approximated as a constant.
 Then   both the energy flux and its   fluctuation  diverge in the $G\to 0$ limit.
 Here the fluctuation originates from the breaking of the dS isometry associated with the static time translation, which induces the uncertainty of time, and also the fluctuation in the value of $H$ at the end of inflation.
 As a result, the inflationary quasi-dS space can be described by Type II$_\infty$ algebra.
 This is different from Type II$_1$ algebra for perfect dS space : since the horizon radius fluctuates by the uncertainty of the observer's energy only,  the entropy of any quantum state cannot exceed that of empty dS space in the Bunch-Davies state.
  In contrast, in Type II$_\infty$ algebra for quasi-dS space, due to the divergent fluctuation of the energy flux,  the trace hence the entropy is not well defined for  the identity describing the maximal entanglement of the Bunch-Davies state, and there is no upper bound on the entropy.

On the other hand, there has been a claim called the `dS swampland conjecture' that dS space is unstable in quantum gravity, which is supported by the distance conjecture and the covariant entropy bound \cite{Obied:2018sgi, Andriot:2018mav, Garg:2018reu, Ooguri:2018wrx}.
Estimation based on the conjecture suggests the much shorter inflationary period given by  $(\epsilon_H^{-1/2} H^{-1})\log (M_{\rm Pl}/H)$, after which $\epsilon_H$ becomes ${\cal O}(1)$ and the background geometry is no longer close to dS space \cite{Seo:2019wsh, Cai:2019dzj}.
Even shorter inflationary period $H^{-1}\log (M_{\rm Pl}/H)$ was conjectured under the name of `trans-Planckian censorship conjecture',  which forbids the horizon crossing of the trans-Planckian modes \cite{Bedroya:2019snp} (see also \cite{Brahma:2019vpl, Brahma:2020zpk, Bedroya:2020rmd, Bedroya:2022tbh}).
In these cases, the energy flux across the horizon, or $\langle \Phi|H_R|\Phi\rangle$ is given by $\epsilon_H^{1/2}[1/(G H)]\log(M_{\rm Pl}/H)$ and $\epsilon_H [1/(G H)]\log(M_{\rm Pl}/H)$, respectively.
Since $\epsilon_H \sim {\cal O}(G)$, the former still diverges but the latter is  of ${\cal O}(1)$ in the $G\to 0$ limit.
In both cases, the fluctuation $\delta H_R$ becomes vanishing in the $G\to 0$ limit as the values in two cases are  estimated as $(\epsilon_H/\pi)^{1/2}\log(M_{\rm Pl}/H)[\log(M_{\rm Pl}/H)+1]$ and $(\epsilon_H^3/\pi)^{1/2}\log(M_{\rm Pl}/H)[\log(M_{\rm Pl}/H)+1]$, respectively.
Hence, the renormalized operator $H'_R=H_R-\langle \Phi|H_R|\Phi\rangle$ is   well defined.    
Then in this limit, we do not need to consider the probability distribution $g(X)$ reflecting the fluctuation in time, and the von Neumann algebra can be defined in the same way as that in the pure dS space with observer, i.e., Type II$_1$ algebra.
 This shows that in addition to the nonzero energy flux across the horizon, the inflationary period plays the crucial role in determining the appropriate von Neumann algebra description of spacetime during the inflation.

%

%


\appendix

\renewcommand{\theequation}{\Alph{section}.\arabic{equation}}

\section{Coordinate systems on de Sitter space}
\label{app:dSdic}

 We list several  coordinate descriptions of dS space which are useful in our discussion. 
 For more complete reviews, see, e.g., \cite{Spradlin:2001pw, Kim:2002uz}.
 A natural description of dS space as seen by a static observer surrounded by the horizon is the static coordinates, in which the metric is written as
 \dis{ds^2=-(1-H^2 r_s^2)dt_s^2+\frac{dr_s^2}{1-H^2 r_s^2}+r_s^2(d\theta^2+\sin^2\theta d\phi^2).\label{eq:staticmetric}}
 From this, one immediately finds that the timelike Killing vector   is just given by $k^a=(\partial_{t_s})^a$ and the horizon is located at $r_s=H^{-1}$.
 In order to see the causal structure, it is convenient to introduce the tortoise coordinate,
 \dis{dr_*=\frac{dr_s}{1-H^2 r_s^2},\quad r_*=\frac{1}{2H}\log\Big(\frac{1+Hr_s}{1-Hr_s}\Big), \label{eq:tortoise}}
 such that the $(t_s, r_*)$ part of the metric is written in the  conformally flat form.
 Then one can define the Eddington-Finkelstein coordinates $u=t_s-r_*$ and $v=t_s+r_*$ as lightcone coordinates.
 For the extension to the region beyond the horizon,  the Kruskal-Szekeres coordinates are useful.
 In the static patch, they are given by  
 \dis{&U =\frac{1}{H}e^{Hu}=\frac{1}{H}e^{Ht_s}\sqrt{\frac{1-Hr_s}{1+Hr_s}\label{eq:UVDef}},
 \\
 &V=-\frac{1}{H}e^{-Hv}=-\frac{1}{H}e^{-Ht_s}\sqrt{\frac{1-Hr_s}{1+Hr_s}},}
  in terms of which the metric is written as
 \dis{ds^2=-\frac{4}{(1-H^2 UV)^2} dUdV+\frac{(1+H^2 UV)^2}{H^2(1-H^2UV)^2}(d\theta^2+\sin^2\theta d\phi^2).\label{eq:UVmetric}}
Using \eqref{eq:UVDef},  the timelike Killing vector in dS is rewritten as $k^a=(\partial_{t_s})^a=H(U\partial_U-V\partial_V)^a$.
 The future (past) horizon is a null hypersurface satisfying $V=0$ ($U=0$), which is normal to $k^a=H U \partial_U$ ($k^a=-H V \partial_V$) or $k_a=-2 HU (dV)_a$ ($k_a=2 HV (dU)_a$).
 Thus, $U$ ($V$) is a natural canonical affine parameter on the future (past) horizon.
 
 Meanwhile, in order to describe the inflationary cosmology, the flat coordinates $(t, r)$ in terms of which the metric is written as
 \dis{ds^2=-dt^2+e^{2Ht}\big[dr^2+r^2(d\theta^2+\sin^2\theta d\phi^2)\big]\label{eq:flatmetric}}
 are useful.
 They are related to the static coordinates by
 \dis{t_s=t-\frac{1}{2H}\log\big(1-H^2 r^2 e^{2Ht}\big),\quad r_s=r e^{Ht},\label{eq:statflat1}}
 which give
 \dis{&\frac{\partial t_s}{\partial t}=\frac{1}{1-H^2 r_s^2},\quad \frac{\partial r_s}{\partial t}= Hr_s,\quad\frac{\partial t_s}{\partial r}=\frac{e^{Ht}Hr_s}{1-H^2r_s^2},\quad \frac{\partial r_s}{\partial r}=e^{Ht},
 \\
&\frac{\partial t}{\partial t_s}=1,\quad \frac{\partial r}{\partial t_s}=-H r,\quad \frac{\partial t}{\partial r_s}=-\frac{Hr_s}{1-H^2 r_s^2 },\quad \frac{\partial r}{\partial r_s}=\frac{e^{-Ht}}{1-H^2 r_s^2 }.\label{eq:statflat2}}
 Then the timelike Killing vector is written as $k^a=(\partial_{t_s})^a=(\partial_t - Hr \partial_r)^a$.
 
\section{Derivation of the density matrix}
\label{app:derden}

Here we briefly sketch how   \eqref{eq:rhoPhi}, the expression for the density matrix $\rho_{\widehat{\Phi}}$  associated with  $|\widehat{\Phi}\rangle$ is obtained, following \cite{Chandrasekaran:2022cip}. 
The relation  \eqref{eq:relmod}, $\langle \Psi|\Delta_{\Phi|\Psi}a|\Psi\rangle=\langle \Phi|a|\Phi\rangle$ indicates that the ${\cal A}_R$ part of $\rho_{\widehat{\Phi}}$  is given by $\Delta_{\Phi|\Psi}$.
Meanwhile, from the facts that the algebra  $\widehat{{\cal A}}_R$ is generated by $\{a, \widehat{q}-(H_0+X)\}$ where $H_0=\beta h_\Psi$ and  that  $h_\Psi-h_{\Phi|\Psi} =h_\Phi-h_{\Psi|\Phi}$ belongs to ${\cal A}_R$, one finds that  the combination
\dis{e^{\beta(-X+\widehat{q})}\Delta_{\Phi|\Psi}=e^{-\beta X+\beta(\widehat{q}-h_\Psi/\beta)+(h_\Psi-h_{\Phi|\Psi})}}
is well factorized into ${\cal A}_R\otimes B(\mathbb{R})\otimes B(\mathbb{R}^+)$.
This motivates us to consider an ansatz
\dis{\rho_{\widehat{\Phi}}=\frac{1}{\beta^2}g(X)^* f(\widehat{q}-h_\Psi/\beta)^*e^{\beta \widehat{q}}\Delta_{\Phi|\Psi}f(\widehat{q}-h_\Psi/\beta)g(X) +{\cal O}(\varepsilon).}
Since we assume that both $f(\widehat{q}-h_\Psi/\beta)$ and $g(X)$ are   slowly varying such that they are taken to be almost constant over $|q-h_\Psi/\beta|<1/\varepsilon$ and $|x|<1/\varepsilon$, respectively,  their commutators  with any other operators are expected to be suppressed by ${\cal O}(\varepsilon)$.
Then $\rho_{\widehat{\Phi}}$ can be written in the form of \eqref{eq:rhoPhi}.

To see $\rho_{\widehat{\Phi}}$ satisfies \eqref{eq:defrhoPhi}, we consider  the operator 
\dis{\widehat{a}=a e^{-i u(\beta (\widehat{q}-X)- h_\Psi)},}
which does not vanish for $\beta|u| <\varepsilon$ as it avoids the  strong oscillation.
Imposing $e^{iuh_\Psi}=1+{\cal O}(\varepsilon)$, its  expectation value with respect to $|\widehat{\Phi}\rangle=|\Phi\rangle\otimes g(X)\otimes f(\widehat{q})$  is given by 
\dis{\langle \widehat{\Phi}|\widehat{a}|\widehat{\Phi}\rangle &=\int_{-\infty}^\infty dx |g(x)|^2 \int_0^\infty dq |f(q)|^2 \langle\Phi|ae^{-i u(\beta  ({q}-x)- h_\Psi)}|\Phi\rangle 
\\
&= \int_{-\infty}^\infty dx |g(x)|^2 \int_0^\infty dq |f(q)|^2e^{-i u\beta  ({q}-x)}\langle\Phi|a|\Phi\rangle+{\cal O}(\varepsilon).}
Here $\langle\Phi|a|\Phi\rangle$ can be replaced by $\langle\Psi|\Delta_{\Phi|\Psi}a|\Psi\rangle$.
Ignoring ${\cal O}(\varepsilon)$ terms and using $h_\Psi|\Psi\rangle=0$, it becomes
\dis{\langle \widehat{\Phi}|\widehat{a}|\widehat{\Phi}\rangle &= \int_{-\infty}^\infty dx \int_0^\infty dq \langle\Psi|\big|g(x+h_\Psi/\beta)\big|^2\big|f(q)\big|^2e^{-i u(\beta ({q}-x)- h_\Psi)}\Delta_{\Phi|\Psi}a|\Psi\rangle
\\
&= \int_{-\infty}^\infty dx \int_0^\infty dq \langle\Psi|\big|g(x+h_\Psi/\beta)\big|^2\big|f(q )\big|^2 \Delta_{\Phi|\Psi}\widehat{a}|\Psi\rangle
\\
&=\int_{-\infty}^\infty \beta dx e^{\beta x} \int_{0}^\infty \beta dq e^{-\beta q }\langle\Psi|\frac{1}{\beta^2}\big| g(x+h_\Psi/\beta) f(q )\big|^2 e^{\beta (-x+q)}\Delta_{\Phi|\Psi}\widehat{a}|\Psi\rangle.}
Matching this with $\int \beta dx e^{\beta x}\langle \Psi_{\rm max}|\rho_{\widehat{\Phi}}\widehat{a}|\Psi_{\rm max}\rangle$, we find that ${\rho}_{\widehat{\Phi}}$ is written as \eqref{eq:rhoPhi}.

\subsection*{Acknowledgements}

This work was supported by the National Research Foundation of Korea (NRF) grant funded by the Korea government (MSIT) (2021R1A4A5031460).

\end{document}